\documentclass[conference]{IEEEtran}

\usepackage{psfrag}
\usepackage{adjustbox}
\usepackage{amssymb}
\usepackage{amsmath}

\newcommand{\be}{\begin{equation}}
\newcommand{\ee}{\end{equation}}
\newcommand{\bea}{\begin{eqnarray}}
\newcommand{\eea}{\end{eqnarray}}

\newcommand{\E}{{\mathbb{E}}}
\newcommand{\A}{{\bf A}}

\title{Asynchronous Performance of Circularly Pulse-Shaped Waveforms for 5G}
\author{\normalsize Ahmad RezazadehReyhani and Behrouz Farhang-Boroujeny 
\\ ECE Department, University of Utah, USA\\
}

\begin{document}
\maketitle

\begin{abstract}\label{sec:abstract}
The fifth generation of wireless networks (5G) necessitates the use of waveforms with loose constraints on synchronization in multiuser scenarios. Also, carrier aggregation, as a way to better utilize the spectrum in 5G, needs a waveform with low out-of-band (OOB) emission. Generalized frequency division multiplexing (GFDM) and circular filter bank multicarrier (C-FBMC) are two candidate waveforms that fulfill these requirements. Both GFDM and C-FBMC operate based on circular convolution, and use cyclic prefix to combat channel response. In this paper, we develop an analytical technique for examining the OOB emission and multiuser interference (MUI) in circularly shaped waveforms, like GFDM and C-FBMC. To stay focused, the study in this paper is limited to C-FBMC modulation. However, the approach we take is trivially extendable to other waveforms as well. We derive equations that quantify OOB emission and MUI. Our analysis allows us to identify the source of OOB emission and MUI. This leads us to quantify the methods proposed by other researchers to decrease OOB emission and MUI. Moreover, we quantify the impact of signal windowing at the transmitter and receiver in reducing OOB emission and MUI, respectively.

\vspace{0.2cm}
keywords: GFDM, C-FBMC, OOB emission, Asynchronous multiuser network, MUI, 5G
\end{abstract}

\section{Introduction}\label{sec:Intro}
Generalized frequency division multiplexing (GFDM), \cite{fettweis2009gfdm}, and circular filter bank multicarrier (C-FBMC), \cite{lin2014advanced}, are two recent candidate waveforms that have been proposed as possible replacements to orthogonal frequency division multiplexing (OFDM) in multiuser environments. OFDM is highly sensitive to time/frequency mismatch and requires strict synchronization. In addition, OFDM generates high out of band (OOB) emission which makes carrier aggregation difficult. 

Both GFDM and C-FBMC are proposed to resolve limitations of OFDM and they operate based on the same principle and closely follow that of OFDM. In particular, a cyclic prefix (CP) is used to absorb channel transient. However, instead of adding a CP to each OFDM symbol, consisting of a cluster of data symbols that are distributed over $N$ subcarriers, a single CP is added to a full packet consisting of MN data symbols that are spanned over $N$ subcarriers and $M$ instants in time. Also, each subcarrier sequence is filtered to confine its respective signal to a limited bandwidth. Moreover, to allow the use of CP, the filtering operations are done through a set of circular convolutions.

Subcarrier filtering results in a lower OOB emission in GFDM/C-FBMC when compared to OFDM. However, due to block structure of transmit signal, their OOB emission may still remain non-negligible. In \cite{michailow2012generalized}, \cite{matthe2014influence} and \cite{michailow2014generalized} it has been recognized that the symbols at packet boundaries are major contributor to high OOB emission. So, they suggested the use of one or more zero-valued guard symbols/subcarriers at GFDM packet time/frequency boundaries to decrease OOB emission. Another method is to use some virtual carriers at edge frequencies with values which effectively cancel OOB emission of the GFDM/C-FBMC signals, \cite{datta2014improved}. These methods clearly lead to some loss of spectral resources and thus make GFDM/C-FBMC inefficient. This goes against the initial intent that the designers of GFDM and C-FBMC had, \cite{fettweis2009gfdm,lin2014advanced}. On the other hand, \cite{matthe2014influence} and \cite{gaspar2015frequency} improved OOB emission by using \mbox{sinc} function or a short pulse as transmitter prototype filter. This method has some side effects, e.g., sensitivity to timing offset. Others have proposed the use of windowing methods to reduce OOB emission, as in filtered OFDM \cite{farhang2011ofdm}. Sample publications that elaborate on this approach are \cite{michailow2014generalized}, \cite{lin2014advanced}, \cite{lin2014multi}, \cite{schellmann2014fbmc} and \cite{abdoli2013weighted}. The subject of multiuser interference (MUI) has recently brought up in \cite{matthe2015asynchronous} and \cite{aminjavaheri2015impact}. The authors of \cite{matthe2015asynchronous} study a GFDM setup in the uplink of a wireless sensor networks (WSN), and \cite{aminjavaheri2015impact} compares the MUI for a few 5G candidate waveforms. The conclusions drawn in all the above papers are mostly based on intuitions and computer simulations.

This paper makes the following contributions. We introduce a novel analysis that allows us to get deeper into the details of the sources that contribute towards OOB emission and MUI in circularly shaped waveforms; specifically, GFDM and C-FBMC. We present a mathematical framework that facilitates our analysis. Our analysis quantifies some of the suggestions made in the previous publications for reducing OOB emission and/or reducing MUI. In addition, we borrow some ideas from the digital subscriber lines literature, \cite{sjoberg1999zipper}, to reduce MUI in GFDM/C-FBMC.

We note that both GFDM and C-FBMC operate based on the same principle. Filtering in each subcarrier band is performed through a circular convolution to allow the use of a CP for combating channel frequency selectivity. However, while GFDM is a non-orthogonal modulation, C-FBMC can be classified as a orthogonal one. The non-orthogonality of GFDM adds some complexities when it comes its study. Such complexities do not exist in C-FBMC, \cite{ahmad2015circularly}. Noting this, in the rest of this paper, we limit our study to C-FBMC, but, we note that all of our findings are extendable to GFDM with some minor changes.

This paper is organized as follows. Section \ref{sec:C-FBMC} presents a brief review of C-FBMC waveform and its respective signal model. Equations that quantify the OOB emission of the waveform are presented in Section \ref{sec:OOB}. In Section \ref{sec:Interference}, we present analytical equations that quantify MUI when users are not synchronized. Section \ref{sec:sim} presents simulation results that provide more insight to MUI under different conditions. Concluding remarks of the paper are made in Section \ref{sec:Conclusion}.

\section{C-FBMC Waveform}\label{sec:C-FBMC}
A C-FBMC waveform, i.e., a single packet carrying a block of data symbols, is constructed as follows. A set of $2MN$ real-valued data symbols, arranged in an $N\times 2M$ matrix $\A$, are built into a waveform for transmission. The alternate elements in a row of $\A$ are in-phase and quadrature parts of QAM symbols, following the offset QAM principles, \cite{farhang2011ofdm}. The elements in each row of $\A$ are transmitted through a subcarrier band. The filtering operation that is used to limit the spectrum of each subcarrier band is performed through a circular convolution, \cite{fettweis2009gfdm,lin2014advanced}. This allows the use of a CP to absorb the channel transient, as done in OFDM.

Here, we present a novel formulation of the waveform construction for C-FBMC that facilitates our analysis in the rest of the paper. This presentation is not intended to introduce any efficient implementation and thus its more complex structure than those reported in the literature, \cite{lin2014multi,farhang2015derivation}, should not be a concern. 

\begin{figure}[tb!]
	\centering
	\includegraphics[width=3.3in]{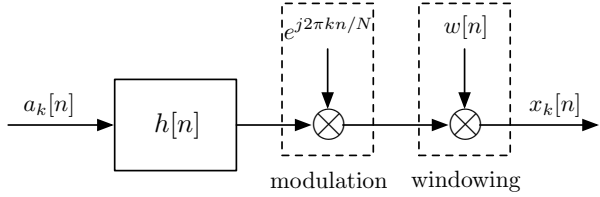}
	\caption{System block diagram for generation of the $k$th subcarrier signal of a circularly shaped waveform.}
	\label{fig:C-FBMCconstruction}	
\end{figure}

To present our formulation, we first concentrate on the contribution of the $k$th row of $\A$ to the generated C-FBMC signal, $x[n]$. Fig.~\ref{fig:C-FBMCconstruction} presents the steps that may be taken to generate the contribution of the $k$th row of $\A$. The result is the output signal $x_k[n]$. The C-FBMC signal, $x[n]$, is obtained by adding the contributions from all the rows of $\A$. 

In Fig.~\ref{fig:C-FBMCconstruction}, the input signal $a_k[n]$ is defined as
\be
a_k[n]=\sum_{l=1}^{2M} j^{k+l}a_{k,l}\delta\left[n-l\frac{N}{2}\right]
\ee
where $a_{k,l}$, for $l=1,2,\cdots,2M$, are the elements of the $k$th row of $\A$, and the additional factor $j^{k+l}$ is to introduce a phase shift of $90^\circ$ among adjacent symbols, following the OQAM multicarrier modulation, \cite{farhang2011ofdm}. The impulse response $h[n]$ is a periodic signal that is constructed by periodically repeating the impulse response of the prototype filter $g[n]$ of C-FBMC waveform. It is assumed that $g[n]$ has a length of $MN$ samples and this is also equal to the length of a period of $h[n]$. Thus, $g[n]$ and $h[n]$ are related as
\be
h[n]=\sum_{i=-\infty}^{\infty} g[n-iMN].
\ee 

The modulator in Fig.~\ref{fig:C-FBMCconstruction} shifts the generated baseband signal to the respective subcarrier band. It is worth noting that the generated signal at the modulator output, by construction, is periodic and has a period of $MN$ samples.
The windowing block truncates one period of this periodic signal plus an additional segment prior to it for the CP part of the packet. Roll-offs may be also added to both sides of $w[n]$ to improve on the OOB emission of the generated C-FBMC signal.

\section{OOB Emission}\label{sec:OOB}
This section is devoted to a study of energy spectral density (ESD) of a C-FBMC signal at the transmitter output. Before we dive into detailed equations, we note that the ESD of $x[n]$ can be written as the summation of contributions from the individual elements $a_{k,l}$ of $\A$, assuming the data symbols $a_{k,l}$ are independent of one another. We also note the contributions of the elements $a_{k,l}$ to the ESD of $x[n]$ differ, depending on their position in the matrix $\A$. Noting these, in the rest of this section, we concentrate on derivation of the ESD of the signal
\be\label{eqn:xkl(n)}
x_{k,l}[n]=j^{k+l}a_{k,l}h\left[n-l\frac{N}{2}\right]e^{j\frac{2\pi k}{N}n}w[n].
\ee
We note that the ESD of $x_{k,l}[n]$ is given by
\be\label{eqn:Ekl(omega)}
E_{k,l}(\omega)=\left|X_{k,l}(\omega)\right|^2
\ee 
where $X_{k,l}(\omega)$ is the discrete-time Fourier transform (DTFT) of $x_{k,l}[n]$.

To facilitate our derivations as well as interpretation of the developed results, we assume that the prototype filter $g[n]$ is the one proposed by Martin and Mirabbasi, \cite{martin1998small, mirabbasi2003overlapped}. This design has been widely accepted in the FBMC community and often referred to as PHYDYAS filter, \cite{bellanger2010fbmc}. A PHYDYAS filter $g[n]$ with the length of $MN$ which is designed to be a square-root Nyquist (N), has the Fourier series expansion
\be
g[n]=\sum_{r=-(M-1)}^{M-1}c_r e^{j\frac{2\pi}{MN}rn}.
\ee
Also, the coefficients $c_r$ are real valued and even symmetric around $r=0$, that is, $c_r=c_{-r}$. 
With the above choice of $g[n]$, the DTFT of the periodic signal $h[n]$ is obtained as
\be\label{eqn:H(omega)}
H(\omega) = \sum_{r=-(M-1)}^{M-1} c_r\delta\left(\omega-\frac{2\pi r}{MN}\right).
\ee

Next, we note that (\ref{eqn:xkl(n)}) implies
\be\label{eqn:Xkl(omega)}
X_{k,l}(\omega)=j^{k+l}a_{k,l} \left(e^{-j\left(\omega-\frac{2\pi k}{N}\right)\frac{lN}{2}}H(\omega-\frac{2\pi k}{N})\right)\star W(\omega)
\ee
where $W(\omega)$ is the DTFT of $w[n]$. Substituting (\ref{eqn:H(omega)}) in (\ref{eqn:Xkl(omega)}) and the result in (\ref{eqn:Ekl(omega)}), we obtain
\be\label{eqn:Ekl(omega)2}
E_{k,l}(\omega) \hspace{-0.05cm}=\hspace{-0.05cm} \sigma_a^2\left|\sum_{r=-(M-1)}^{M-1} \hspace{-0.5cm} c_r e^{-j\pi \frac{lr}{M}}W \hspace{-0.1cm} \left(\omega-\frac{2\pi (r+kM)}{MN}\right)\right|^2
\ee 
where $\sigma_a^2=\E[|a_{k,l}|^2]$, and $\E[\cdot]$ denotes expectation.

\begin{figure}[tb!]
	\centering
	\includegraphics[width=3.3in]{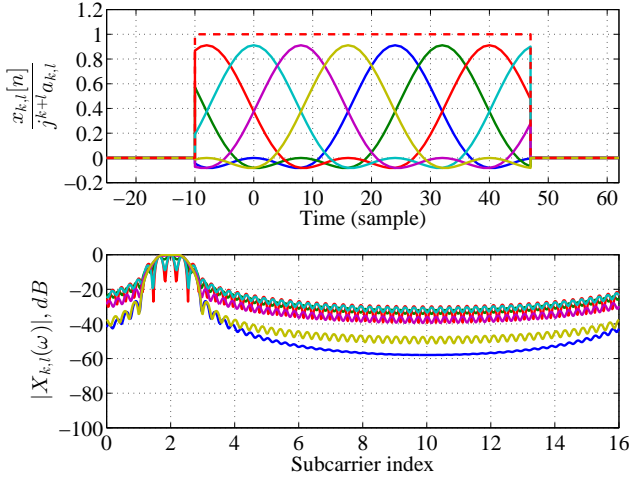}
	\caption{CP appended circular pulses and their respective amplitude responses in a C-FBMC data block and in a single subcarrier band. Here a rectangular window is used to time limit the generated block.}
	\label{fig:GFDMPulsesCP}	
\end{figure}

\begin{figure}[tb!]
	\centering
	\includegraphics[width=3.3in]{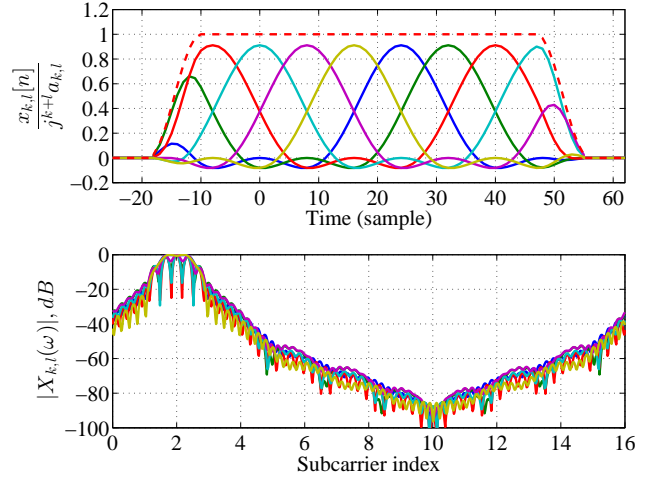}
	\caption{CP appended circular pulses and their respective amplitude responses in a C-FBMC data block and in a single subcarrier band. Here a raised-cosine window is used to time limit the generated block.}
	\label{fig:GFDMPulsesWin}	
\end{figure}

Fig.~\ref{fig:GFDMPulsesCP} presents the CP appended circular pulse shapes for all values of the parameter $l$ in a C-FBMC data block and their respective amplitude responses in a single subcarrier. Here, there are $N=16$ subcarriers and there are $2M=6$ real symbols across time. The length of CP is equal to 10 and the generated data block is windowed using a rectangular window. The rectangular window is the dashed-line red plot. The selected subcarrier is the subcarrier number 2. The results show that while the pulse shapes near the center of the block have a well contained spectrum within the respective subcarrier band, the pulse shapes that are closer to the edges of the block have a significant OOB emission. Such OOB emissions are clearly due to the sharp transitions at the beginning and the end of the block. One may also note that such OOB emissions can be reduced significantly by extending the length of the block and introducing a roll-off to its edges as demonstrated in Fig.~\ref{fig:GFDMPulsesWin}. This solution which follows the approach of filtered OFDM, \cite{farhang2011ofdm}, has also been mentioned in \cite{michailow2014generalized,lin2014advanced,lin2014multi,schellmann2014fbmc,abdoli2013weighted}. Others have suggested nulling the symbols that are near the edges of the block, \cite{michailow2012generalized,datta2014improved,matthe2014influence,michailow2014generalized}. This latter solution is not a desirable one, as it incurs a significant loss in spectral efficiency.

It is also instructive to pay some attention to the mathematical details in (\ref{eqn:Ekl(omega)2}) and relate our observation there to the energy spectral densities in Figs.~\ref{fig:GFDMPulsesCP} and \ref{fig:GFDMPulsesWin}. When $w[n]$ is a rectangular window (Figs.~\ref{fig:GFDMPulsesCP}), $W(\omega)$ is a sinc-like function of $\omega$, clearly, with side-lobes that are relatively large. For $l=M$, the choice of the coefficients $c_r$ has been made such that the side-lobes of the terms under the summation in (\ref{eqn:Ekl(omega)2}) cancel each others and, thus, lead to a minimal OOB emission. As $l$ deviates from $M$, the added phase rotations in the coefficients $c_r e^{-j\pi \frac{lr}{M}}$ in (\ref{eqn:Ekl(omega)2}) introduces some imbalance in the said side-lobes cancellation and as a result the OBB emission deteriorates. On the other hand, the introduction of roll-offs to the ends of the window function $w[n]$ reduces the side-lobes of $W(\omega)$ and as a result OBB emission will have less dependency on the parameter $l$, i.e., the symbol position along the time.

\section{Multiuser Interference}\label{sec:Interference}
In the up-link of a multiuser network, signals from different users may reach the base station out of sync. In this section, we quantify MUI by looking at interference between an arbitrary pair of data symbols $a_{k,l}$ and $a_{p,m}$ that are transmitted asynchronously from two mobile terminals. 

We consider the case where the receiver is synchronized to detect $a_{p,m}$ and study the interference introduced by $a_{k,l}$. We assume the signal that carries $a_{k,l}$ is received with a time offset of $\Delta n$ and a frequency offset of $\Delta f$. The associated received signal is thus given by 
\be\label{eqn:y(n)}
y[n] = x_{k,l}[n-\Delta n]e^{j2\pi\Delta f n}.
\ee

The receiver selects a block of the received signal with a window $v[n]$, and down converts the $p$th subcarrier to the baseband. Accordingly, the portion of the down converted signal that arises from $a_{k,l}$ is obtained as
\be\label{eqn:r(n)}
z[n] = y[n] v[n] e^{-j\frac{2\pi}{N}pn}.
\ee

\begin{figure}[tb!]
	\psfrag{C}[][]{{$\scriptstyle N_{cp}$}}
	\psfrag{N}[][]{{$\scriptstyle MN$}}
	\centering
	\includegraphics[width=3.3in]{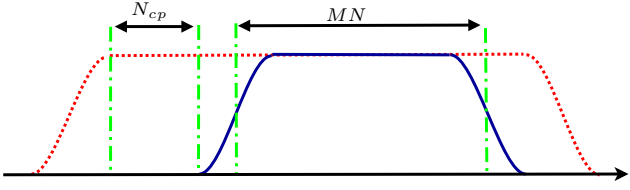}
	\caption{Transmitter window, $w[n]$(red line), and receiver window, $v[n]$(blue line)}
	\label{fig:Windowing}	
\end{figure}

Next, we define $I_{p,m}^{k,l}$ as the leakage gain between the transmit data symbol $a_{k,l}$ and the receiver output that delivers an estimate of $a_{p,m}$. This leakage gain can be calculated by using the following formula.
\begin{align}
I_{p,m}^{k,l} &= \frac{1}{a_{k,l}}\Re{\left\{j^{-\left(p+m\right)} z[n] \star h[n] \bigg|_{n=m\frac{N}{2}} \right\}}\nonumber\\
 &= \frac{1}{a_{k,l}}\Re{ \left\{j^{-\left(p+m\right)} \frac{1}{2\pi}\int_{0}^{2\pi} \hspace{-0.2cm} Z(\omega)H(\omega)e^{j\omega m\frac{N}{2}}d\omega\right\} }\label{eqn:apm1}
\end{align}
where $Z(\omega)$ is the DTFT of $z[n]$. Substituting (\ref{eqn:H(omega)}) in (\ref{eqn:apm1}) and rearranging the results, we obtain
\be\label{eqn:apm2}
I_{p,m}^{k,l}= \frac{1}{a_{k,l}}\Re{ \left\{j^{-\left(p+m\right)} \hspace{-0.3cm} \sum_{r=-(M-1)}^{M-1} \hspace{-0.2cm} c_r Z\left(\frac{2\pi r}{MN}\right)e^{j\frac{\pi mr}{M}}\right\} }.
\ee

To develop some insight into the values of $I_{p,m}^{k,l}$ under different conditions, 
we substitute (\ref{eqn:xkl(n)}) in (\ref{eqn:y(n)}) and the result in (\ref{eqn:r(n)}) to obtain
\be
z[n]\hspace{-0.1cm}=\hspace{-0.1cm}j^{k+l}\hspace{-0.05cm}a_{k,l}h\hspace{-0.1cm}\left[n\hspace{-0.1cm}-\Delta n\hspace{-0.1cm}-l\frac{N}{2}\right] e^{-j\frac{2\pi k}{N}\Delta n} e^{j2\pi\left(\frac{k-p}{N}+\Delta f\right)n}u[n] 
\ee
where $u[n]$ is the combined window function $u[n]=w[n-\Delta n]v[n]$.
Applying DTFT to both sides of this result, we arrive at the result that is presented in equation (\ref{eqn:R(omega)}) at the top of the next page. Also, substituting (\ref{eqn:R(omega)}) in (\ref{eqn:apm2}), we obtain equation (\ref{eqn:apm3}) that is also presented at the top of the next page. In (\ref{eqn:R(omega)}), $U(\cdot)$ is the DTFT of the combined window function $u[n]$.
\begin{figure*}[tb]
	\be\label{eqn:R(omega)}
	R(\omega) = a_{k,l} j^{k+l} e^{-j\frac{2\pi}{MN}k\Delta n} \hspace{-0.3cm} \sum_{r'=-(M-1)}^{M-1} \hspace{-0.2cm} c_{r'} e^{-j\frac{2\pi r'}{MN}\left(\Delta n+\frac{lN}{2}\right)} U\left(\omega- \frac{2\pi}{N}(k-p)+2\pi\Delta f-\frac{2\pi}{MN}r'\right)
	\ee
	\be\label{eqn:apm3}
	I_{p,m}^{k,l}= \Re{ \left\{j^{k-p+l-m} e^{-j\frac{2\pi}{MN}k\Delta n} \hspace{-0.3cm} \sum_{r=-(M-1)}^{M-1} \hspace{-0.3cm} c_r e^{j\frac{\pi mr}{M}} \hspace{-0.3cm} \sum_{r'=-(M-1)}^{M-1} \hspace{-0.3cm} c_{r'} e^{-j\frac{2\pi r'}{MN}\left(\Delta n+\frac{lN}{2}\right)} U\left(\frac{2\pi}{MN}(r-r'-(k-p)M)+2\pi\Delta f\right) \right\} }
	\ee
	\hrulefill
\end{figure*}

 Fig.~\ref{fig:Windowing} shows the transmitter window function $w[n]$ and the receiver window function $v[n]$ and their respective positions. We note that in a synchronized case $u[n]=v[n]$. The roll-offs of the window functions, shown in Fig.~\ref{fig:Windowing}, as noted earlier, helps in reducing OOB emission and MUI. Also, an important property of the receiver window $v[n]$ that will become useful in the rest of our discussion is the following:
\be\label{eqn:property_of_V}
V\left(\frac{2\pi}{MN}r\right)=\left\{\begin{array}{ll}
1, & r=0\\
0, & r\neq 0.\end{array}\right.
\ee

In the sequel, we dig into the specific details of (\ref{eqn:apm3}) under various conditions.

\subsection{Fully synchronized}
In a fully synchronized C-FBMC network, by design, there will be no inter-carrier interference and, thus, MUI is avoided. In other words, $I_{p,m}^{k,l} =0$, for $(p,m)\neq (k,l)$. Here, we present a proof of this fact by looking into details of (\ref{eqn:apm3}) for the case where $\Delta n=\Delta f=0$. This proof/presentation, although may seem unnecessary, plays the roll of an instructive introduction to the rest of our study in this paper. 

We recall that when $\Delta n=\Delta f=0$, $u[n]=v[n]$ and thus the property (\ref{eqn:property_of_V}) will be also applicable to $U(\omega)$ and may be used to simplify (\ref{eqn:apm3}). Such simplification leads to 
\be\label{eqn:full}
I_{p,m}^{k,l} = \Re{\left\{j^{k-p+l-m} \hspace{-0.45cm} \sum_{r=-(M-1)}^{M-1} \hspace{-0.45cm} c_r c_{r-(k-p)M} e^{j\frac{\pi r}{M}(m-l)}\right\}}.
\ee

In the case where the subcarriers $k$ and $p$ are non-adjacent to each other, $|k-p| \ge 2$ and under this condition the set of coefficients $c_r$ and $c_{r-(k-p)M}$, for $-(M-1)\le r\le M-1$ are non-overlapping. This, in turn, implies that all the terms under the summation in (\ref{eqn:full}) are zero, hence, $I_{p,m}^{k,l} = 0$.

In the case where the subcarriers $k$ and $p$ are adjacent, $k-p=\pm 1$. For this case, we study (\ref{eqn:full}) for two scenarios where $m-l$ is odd and when it is even. 

When $m-l$ is an odd number, $j^{k-p+l-m}=\pm 1$ and (\ref{eqn:full}) simplifies to 
\be\label{eqn:full2}
I_{p,m}^{k,l} = \pm \sum_{r=-(M-1)}^{M-1} \hspace{-0.2cm} c_r c_{r\pm M} \cos \left(\frac{\pi r}{M}(m-l)\right).
\ee
Next, we introduce a change of variable $r$ to $r\mp\frac{M}{2}$ to rearrange (\ref{eqn:full2}) as
\begin{align}\label{eqn:full3}
I_{p,m}^{k,l} &=\pm \sum_{r=-\frac{M}{2}+1}^{\frac{M}{2}-1} c_{r+\frac{M}{2}}c_{r-\frac{M}{2}} \sin\left(\frac{\pi r}{M}(m-l)\right)\nonumber\\
&= 0.
\end{align}
Here, the second identity follows since the expression under summation in the first line of (\ref{eqn:full3}) is an odd series in $r$. 

When $m-l$ is an even number, $j^{k-p+l-m}=\pm j$ and (\ref{eqn:full}) simplifies to 
\begin{align}\label{eqn:full4}
I_{p,m}^{k,l} &= \pm \sum_{r=-(M-1)}^{M-1} \hspace{-0.2cm} c_r c_{r\pm M} \sin \left(\frac{\pi r}{M}(m-l)\right)\nonumber\\
&=0.
\end{align}
Here, the second identity follows since the expression under summation can be made an odd series in $r$ through a change of variable $r$ to $r\mp\frac{M}{2}$.

To summarize, the key factors which guarantee interference free operation of C-FBMC in a synchronous scenario are
\begin{itemize}
	\item A phase toggle of $\pi/2$ between adjacent time/frequency data symbols.
	\item Design of a prototype filter such that DTFT coefficients $c_r$ are non-zero only for $-M+1\le r \le M-1$ and are even symmetric around $r=0$.
\end{itemize} 
As we will see below, both timing offset and frequency offset ruins these properties, hence, MUI will be unavoidable in asynchronous networks. Windowing method greatly helps in reducing MUI.

\subsection{Timing offset only}
When $\Delta f=0$, but $\Delta n\ne 0$, two scenarios can happen. First, the window $v[n]$ falls over the flat part of the transmitter window, $w[n]$. This can be due to the presence of a sufficiently long CP and/or a cyclic suffix (CS). In this scenario, $u[n]=v[n]$ and (\ref{eqn:apm3}) reduces to
\begin{align}\label{eqn:full5}
I_{p,m}^{k,l} \hspace{-0.05cm} = \hspace{-0.05cm} \Re{\left\{j^{k-p+l-m} e^{-j\frac{2\pi k}{MN}\Delta n} \hspace{-0.5cm} \sum_{r=-(M-1)}^{M-1} \hspace{-0.5cm} c_r c_{r-(k-p)M} e^{j\frac{\pi r}{M}(m-l)}\right\}}.
\end{align}
This is similar to (\ref{eqn:full}) with the addition of the phase factor $e^{-j\frac{2\pi}{MN}k\Delta n}$. The presence of this phase factor has no impact on the leakage gain $I_{p,m}^{k,l}$ being equal to zero when $|k-p|\ge 2$. However, it introduces some leakage when $k-p=\pm 1$. In the latter case, one finds that when $m-l$ is an odd number, (\ref{eqn:full5}) simplifies to 
\be\label{eqn:TO1}
I_{p,m}^{k,l} = \pm \hspace{-0.5cm} \sum_{r=-(M-1)}^{M-1} \hspace{-0.4cm} c_r c_{r\pm M} \cos \left(\frac{\pi r}{M}(m-l) - \frac{2\pi k}{MN}\Delta n \right)
\ee
and when $m-l$ is an even number, (\ref{eqn:full5}) simplifies to 
\be\label{eqn:TO2}
I_{p,m}^{k,l} = \pm \hspace{-0.5cm} \sum_{r=-(M-1)}^{M-1} \hspace{-0.4cm} c_r c_{r\pm M} \sin \left(\frac{\pi r}{M}(m-l) - \frac{2\pi k}{MN}\Delta n\right).
\ee 
With presence of phase $- \frac{2\pi k}{MN}\Delta n$, expression inside the summations in (\ref{eqn:TO1}) and (\ref{eqn:TO2}) no longer hold the odd symmetry property that was mentioned above, and thus $I_{p,m}^{k,l}$ may not be zero. Hence, MUI will be present.

The second scenario is when the window $v[n]$ does not fall over the flat part of $w[n]$. In this scenario, $u[n]\ne v[n]$ and, thus, $U(\omega)$ no longer holds the {\em clean} property that is stated by (\ref{eqn:property_of_V}). The samples of $U(\omega)$ that appear under the summation in (\ref{eqn:full}) are all non-zero. This results in non-zero values for $I_{p,m}^{k,l}$ for all pairs of $(p,m)$ and $(k,l)$. The addition of proper roll-offs to both $w[n]$ and $v[n]$ helps in moderating the power leakage, but cannot remove it completely. More research into the details of the leakage gain $I_{p,m}^{k,l}$ is beyond the scope of this paper. It will be reported in our future works.

\subsection{Frequency offset only}
The presence of a frequency offset $\Delta f$ shifts the spectrum of the received signal. So, samples of $U(\omega)$ in (\ref{eqn:apm3}) are taken at frequencies other than multiple integers of $\frac{1}{MN}$. All these samples are non-zero. Thus, (\ref{eqn:apm3}) results in a non-zero value for $I_{p,m}^{k,l}$. Extending the window function $v[n]$ with proper roll-off decreases OOB emission of $U(\omega)$. As a result, samples of $U(\omega)$ in (\ref{eqn:apm3}) have smaller values, and the resulting interference gains become smaller. This method reduces the interference gains but their values still remain non-zero.

\section{Simulation Results}\label{sec:sim}
The interference gain $I_{p,m}^{k,l}$ varies with the position of the asynchronous data symbol, $(k,l)$, the position of the estimated data symbol, $(p,m)$, the time offset, $\Delta n$, and the frequency offset, $\Delta f$. In this section, we present a few case studies to gain more insight to the analytical results that were presented in the previous section. 

We simulate a C-FBMC system with $N=16$ subcarriers and $2M=16$ real-valued symbols along the time. The length of CP is set equal to $8$. The channel is assumed to be an ideal one. We present color map plots of $I_{p,m}^{k,l}$ for three choices of $(k,l)=(8,1), (8,3),$ and $(8,7)$ and three set of the time and frequency offsets $(\Delta n, \Delta f)=(4,0),(-4,0),$ and $(0,0.05/N)$. For the cases where roll-off windows are applied, the packet length is extended as indicated in Fig.~\ref{fig:Windowing}. The roll-off lengths at both the transmitter and receiver are set equal to 8.

\begin{figure}[t]
	\centering
	\text{\tiny $~~(\Delta n, \Delta f)\hspace{-0.08cm}=\hspace{-0.08cm}(4,0)~~~~~~~~~~(\Delta n, \Delta f)\hspace{-0.08cm}=\hspace{-0.08cm}(-4,0)~~~~~~~~(\Delta n, \Delta f)\hspace{-0.08cm}=\hspace{-0.08cm}(0,\frac{0.05}{N})$}
	\psfrag{Subcarrier Index}[][]{{\small Subcarrier Index}}
	\psfrag{Time Index}[][]{{\small Time Index}}
	\includegraphics[scale=0.43]{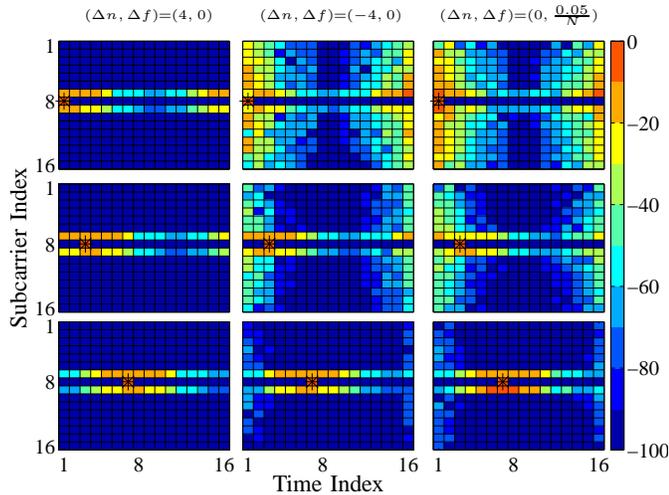}
	\caption{Interference gain of an asynchronous data symbol (shown by *) to other synchronous data symbols. Receiver window $v[n]$ is a rectangular window.}
	\label{fig:Leakage}
\end{figure}

\begin{figure}[t]
	\centering
	\text{\tiny $~~(\Delta n, \Delta f)\hspace{-0.08cm}=\hspace{-0.08cm}(4,0)~~~~~~~~~~(\Delta n, \Delta f)\hspace{-0.08cm}=\hspace{-0.08cm}(-4,0)~~~~~~~~(\Delta n, \Delta f)\hspace{-0.08cm}=\hspace{-0.08cm}(0,\frac{0.05}{N})$}
	\psfrag{Subcarrier Index}[][]{{\small Subcarrier Index}}
	\psfrag{Time Index}[][]{{\small Time Index}}
	\includegraphics[scale=0.43]{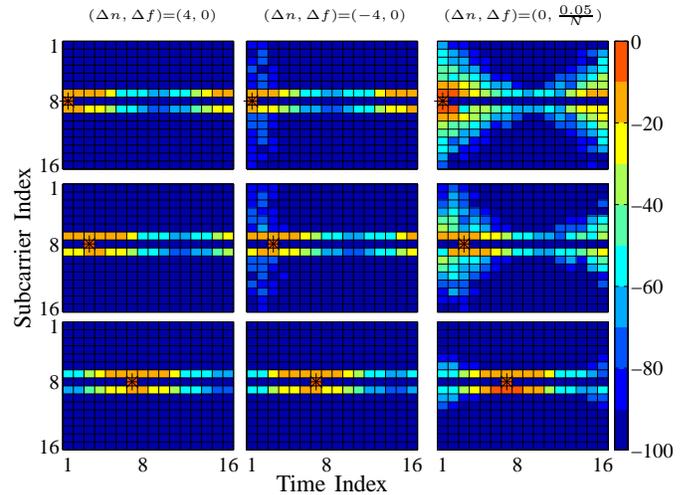}\\[0.5mm]
	
	\caption{Interference gain of an asynchronous data symbol (shown by *) to other synchronous data symbols. Receiver window $v[n]$ is a raised cosine window.}
	\label{fig:LeakageWin}
\end{figure}

Fig.~\ref{fig:Leakage} presents the color map plots of $I_{p,m}^{k,l}$ for the case where rectangular windows are used at both the transmitter and receiver. The same set of color map plots are repeated in Fig.~\ref{fig:LeakageWin} when windows with raise-cosine roll-offs have been used at both the transmitter and receiver. The following observations (that all match our theoretical analysis in the previous section) are made from the color maps:
\begin{itemize}
\item
The asynchronous data symbols which are closer to the center of packet generate less interference.
\item
For $(\Delta n, \Delta f)=(4,0)$, only data symbols in the adjacent subcarriers are subject to interference. This is the case where the selected window of the asynchronous user falls over the flat part of its transmit window. The results are very different when $(\Delta n, \Delta f)=(-4,0)$.
\item
The use of windows with smooth corners has a significant impact on the level of leakage/interference. In particular, for the case where $(k,l)=(8,1)$ and $(\Delta n,\Delta f)=(-4,0)$ or $(0,0.05/16)$, in Fig.~\ref{fig:Leakage} we see a lot of orange and yellow color boxes, indicating leakage gains of $-10$ to $-25$~dB. These mostly drop below $-40$~dB in Fig.~\ref{fig:LeakageWin}.
\end{itemize}


\section{Conclusion}\label{sec:Conclusion}
In this paper, we developed a mathematical framework for analysis of circularly shaped waveforms, such as GFDM and C-FBMC. Our analysis allowed us to identify the source of OOB emission and MUI and to quantify the methods proposed by other researchers for decreasing these undesirable effects. Moreover, our analysis led us to quantify the impact of signal windowing at the transmitter, for reducing OOB emission, and at the receiver, to control MUI.


\bibliographystyle{IEEEtran}
\bibliography{ICC2016}

\end{document}